\documentclass[aps, showpacs, showkeys,nofootinbib,floatfix]{revtex4}

\usepackage{amssymb}
\usepackage{amsmath}
\usepackage{graphicx}

\begin{document}

\title{Observational constraints on unified dark matter including Hubble parameter data}

\author{Kai Liao, Shuo Cao, Jun Wang, Xiaolong Gong and Zong-Hong Zhu}
 \email{zhuzh@bnu.edu.cn}
\affiliation{Department of Astronomy, Beijing Normal University,
Beijing 100875, China}

\begin{abstract}
We constrain a unified dark matter (UDM) model from the latest observational data.
This model assumes that the dark sector is degenerate. Dark energy and dark matter are the same
component. It can be described by an affine equation of state $P_X= p_0 +\alpha \rho_X$.
Our data set contains the newly revised $H(z)$ data, type Ia supernovae (SNe Ia) from Union2 set,
baryonic acoustic oscillation (BAO) observation from the spectroscopic Sloan Digital Sky
Survey (SDSS) data release 7 (DR7) galaxy sample, as well as the cosmic microwave
background (CMB) observation from the 7-year Wilkinson Microwave Anisotropy Probe (WMAP7) results.
By using the Markov Chain Monte Carlo (MCMC) method, we obtain the results in a flat universe:
$\Omega_\Lambda$=$0.719_{-0.0305}^{+0.0264}(1\sigma)_{-0.0458}^{+0.0380}(2\sigma)$,
$\alpha$=$1.72_{-4.79}^{+3.92}(1\sigma)_{-7.30}^{+5.47}(2\sigma)(\times10^{-3})$,
$\Omega_bh^2$=$0.0226_{-0.0011}^{+0.0011}(1\sigma)_{-0.0015}^{+0.0016}(2\sigma)$.
Moreover, when
considering a non-flat universe,
$\Omega_\Lambda$=$0.722_{-0.0447}^{+0.0362}(1\sigma)_{-0.0634}^{+0.0479}(2\sigma)$,
$\alpha$=$0.242_{-0.775}^{+0.787}(1\sigma)_{-1.03}^{+1.10}(2\sigma)(\times10^{-2})$,
$\Omega_bh^2$=$0.0227_{-0.0014}^{+0.0015}(1\sigma)_{-0.0018}^{+0.0021}(2\sigma)$,
$\Omega_k$=$-0.194_{-1.85}^{+2.02}(1\sigma)_{-2.57}^{+2.75}(2\sigma)(\times10^{-2})$.
These give a more stringent results than before. We also give the results from other combinations
of these data for comparison.
The observational Hubble parameter data can give a more stringent constraint than SNe Ia.
From the constraint results, we can see the parameters
$\alpha$ and $\Omega_k$ are very close to zero, which means a flat universe is strongly supported
and the speed of sound of the dark sector seems to be zero.

\end{abstract}


\maketitle

\section{$\text{INTRODUCTION}$}
Cosmic acceleration is one of the most striking discoveries in modern cosmology \cite{acceleration}. Many works have been done
to explain this unexpected phenomenon. There are different mechanisms for the universe acceleration. The most popular idea is assuming
a new component with negative pressure known as dark energy. Other mechanisms contain various modifications of gravitation theory \cite{MGR}
and violation of cosmological principle \cite{LTB}.
There are all
kinds of dark energy models including quintessence \cite{quintessence}, holographic dark energy \cite{holographic}, quintom \cite{quintom}, phantom \cite{phantom}, generalized Chaplygin
gas model \cite{GCG} and
so on. Among these, the simplest and most successful model is $\Lambda$CDM model \cite{LCDM} while it has to face some theoretical problems.
For example, the "coincidence" problem and the "fine-tuning" problem \cite{pr}.
From this standard model, the flat universe consists of $\sim 4\%$ baryons,  $\sim23\%$ dark matter, $\sim 73\%$ dark energy \cite{standard}.
However, we still know little about dark matter and dark energy \cite{unknown}. All we can feel about them are their gravitational effects. Observationally, they are degenerate. Based on these, a unified dark matter model was proposed which assumes that dark matter and
dark energy are from a same dark component \cite{UDM}. Since the equation of state of the unified dark component is unknown, we assume that it has a constant speed of sound $c_s^2=\alpha$ which is equivalent to the affine form \cite{constraint}.
In order to break the degeneracies, different kinds of observational data are used to constrain cosmological model, among which
SNe Ia, CMB, and BAO are usually used to constrain cosmological parameters \cite{observation}. Recently, other observations are included. For example,
the gamma-ray burst \cite{GRB}, cluster gas mass fraction \cite{fgas}, lensing \cite{lensing} and so on. But these processes need to integrate the Hubble parameter to
obtain the distance scale. The more information about $H(z)$ can't be embodied. Therefore, it is meaningful to study the $H(z)$ data
directly. The expression of the Hubble parameter can be written in this form
\begin{equation}
H(z)=-\frac{1}{1+z}\frac{dz}{dt}\,,
\end{equation}
which depends on the differential age as a function of redshift. Based on Jimenez et al. \cite{Jimenez}, Simon et al. \cite{Simon} used the age of evolving
galaxies and got nine $H(z)$ data. Many works based on these nine data was done \cite{nine}. Recently, Stern et al. \cite{Stern} revised these data at 11
redshifts from the differential ages of red-envelope galaxies. Moreover, Gazta\~{n}aga et al. \cite{three} took the BAO scale as a standard ruler in the radial
direction, obtained three more additional data.

In this paper, we use the Markov chain Monte Carlo (MCMC) method to give constraints on the unified dark matter model from the latest
data including the observational Hubble parameter data. In section 2, we give a brief introduction of the UDM model. In section 3, we
introduce the observational data set we use. The full parameter space using different combinations of data is exhibited in section 4.
Finally, we give a conclusion in section 5.

\section{$\text{UNIFIED DARK MATTER WITH CONSTANT SPEED OF SOUND}$}
In this section, we give a brief introduction to unified dark matter model. For details of this model, please refer to \cite{UDM}.

Considering the universe is described by the Friedmann-Robertson-Walker metric and assuming a single UDM besides baryons and radiation,
the Friedmann equation is expressed as
\begin{equation}
H(a)=H_0E(a),
\end{equation}
where $H_0$ represents the Hubble constant today, and
\begin{equation}
\label{einstein1}
E(a)^2=\Omega_r(a)+\Omega_b(a)+\Omega_X(a)+\Omega_k(a),
\end{equation}
where $\Omega_{i}(a)=\rho_{i}(a)/\rho_{c}$, $\rho_{c}$ is the critical density today. The subscripts r, b, k and X represent for radiation, baryons, curvature and UDM component respectively.
Each component satisfies the energy-momentum conservation equation
\begin{equation}
\label{einstein2}
\dot{\rho}_i=-3H(\rho_i+P_i).
\end{equation}

For the radiation and baryons, the equations of state are $1/3$ and $0$, respectively. The conservation equation gives
$\Omega_b(a)\propto a^{-3}$ and $\Omega_r(a)\propto a^{-4}$.

For the UDM, it is considered as an approximation to an barotropic fluid. Since the equation of state $P_X=P_X(\rho_X)$ is unknown, we
assume a constant speed of sound
\begin{equation}
dP_X/d\rho_X\simeq \alpha,
\end{equation}
which leads to the 2-parameter affine form
\begin{equation}
\label{EOSaffine}
P_X\simeq p_0+\alpha \rho_X.
\end{equation}

Combined with the conservation equation, we can regard this component as
an effective cosmological constant plus a generalized dark matter which has
a constant equation of state $\alpha$. The density evolves as
\begin{equation}
\label{rhodia}
\Omega_X(a)=\Omega_\Lambda+(\tilde{\Omega}_m)a^{-3(1+\alpha)},
\end{equation}
where $\tilde{\Omega}_m=\Omega_{Xo}-\Omega_\Lambda$ stands for the generalized dark matter, and
$\Omega_\Lambda$ stands for the effective cosmological constant component, $\dot{\rho}_\Lambda=0$.

The Friedmann equation now can be expressed as
\begin{equation}
\label{einstein1}
E(a)^2=\Omega_{r}a^{-4}+\Omega_{b}a^{-3}+\Omega_\Lambda+(1-\Omega_{b}-\Omega_{r}-\Omega_\Lambda)a^{-3(1+\alpha)}+\Omega_ka^{-2},
\end{equation}
where $\Omega_{r}$, $\Omega_{b}$ and $\Omega_{k}$ represent dimensionless radiation, baryon matter and curvature today. $\Omega_\Lambda$ and $\alpha$ are the parameters we introduced. \textbf{We notice that when $\alpha=0$, the UDM model are equivalent to $\Lambda$CDM. In UDM model, we can equivalently think that the generalized dark matter may has a non-zero equation of state. Naturally, because the property of the effective cosmological constant component is the same as the one in $\Lambda$CDM, when the property of generalized dark matter is equivalent to the usual dark matter in $\Lambda$CDM, UDM model totally becomes $\Lambda$CDM.}

\textbf{The speed of sound characterizes the perturbations to the dark matter density and pressure which are related to the microphysics. Since we know litter about dark sector in our universe, the equation of state is introduced to describe the property of dark sector. However, it can not give us the information about the microphysics and other properties. The speed of sound can give information about the internal degrees of freedom. For example, quintessence has $c_s=1$. If we can confirm the sound speed that is below the speed of light, we can get further degrees beyond a canonical, minimally coupled scalar field. If the sound speed is small, the perturbations can be detected on a proper scale. This perturbations will change the gravitational potential, thus affecting the geodesic of photons. As a result, the cosmic microwave background will be changed through the Integrated Sachs-Wolfe (ISW) effect. In order to show the differences between UDM mdoel and $\Lambda$CDM which corresponds to $\alpha=0$, we give the CMB temperature angular power spectrum in Fig. 7. Our calculations are based on CAMB \cite{camb}.}

\section{$\text{CURRENT OBSERVATIONAL DATA}$}
Now, we introduce the data set using to constrain the UDM model.

\subsection{Observational Hubble parameter data}
The observational Hubble parameter as a function of redshift has been used to constrain cosmological parameters \cite{hzre}.
For the observational data, we choose 11 data from \cite{Stern}. These data were obtained through the measurement of differential ages
of red-envelope galaxies known as "differential age method". The aging of stars can be regarded as an indicator of the aging of the universe.
The spectra of stars can be converted to the information of their ages, as the evolutions of stars are well-known. Since we can not observe stars
one by one at cosmological scales, we take the spectra of galaxies which contain relatively uniform star population. The surveys including the
Gemini Deep Deep Survey \cite{GDDS}, VIMOS-VLT Deep Survey and SDSS \cite{Stern} are related to obtaining data. We consider the red-envelope galaxies which are
massive with fairly homogeneous star population.
 Moreover, the data can be obtained from the BAO scale as a standard ruler in the
radial direction known as "Peak Method" \cite{three}.
Though this method, we have three additional data \cite{three}: $H(z=0.24)=79.69\pm2.32,
H(z=0.34)=83.8\pm2.96,$ and $H(z=0.43)=86.45\pm3.27$, which are independent. Plus the measurement of Hubble constant \cite{h0}, we totally have 15
data now.
The $\chi^2$ value of the $H(z)$ data can be expressed as
\begin{equation}
\label{chi2H}
\chi^2_H=\sum_{i=1}^{15}\frac{[H(z_i)-H_{obs}(z_i)]^2}{\sigma_{i}^2},
\end{equation}
where $\sigma_{i}$ is the $1\sigma$ uncertainty of the observational
$H(z)$ data.

\subsection{Type Ia supernovae}
Since the SNe Ia fist revealed the acceleration of the universe, it has been a well-established method for probing dark energy.

Recently, the Supernova Cosmology Project (SCP) released the Union2 data set including 557 samples \cite{Union2}. It covers a
redshift region 0-1.4. These data have been widely
used to investigate dark energy models \cite{use}. The distance modules is given by
\begin{equation}
\mu_{theory}(z)=5\log_{10}[d_{L}(z)]+\mu_{0},
\end{equation}
where $d_{L}(z)$ is the luminosity distance. It is defined as
\begin{equation}
d_L(z)=(1+z)r(z),\quad r(z)=\frac{c}{H_0\sqrt{|\Omega_{k}|}}{\rm
sinn}\left[\sqrt{|\Omega_{k}|}\int^z_0\frac{dz'}{E(z')}\right].
\end{equation}

The marginalized nuisance parameter \cite{marginalized} for
$\chi^2$ is expressed as
\begin{equation}
\label{chi2SN} \chi^2=A-\frac{B^2}{C}+\ln\left(\frac{C}{2\pi}\right),
\end{equation}
where $A=\sum_i^{557}{(\mu^{\rm data}-\mu^{\rm
theory})^2}/{\sigma^2_{i}}$, $B=\sum_i^{557}{(\mu^{\rm
data}-\mu^{\rm theory})}/{\sigma^2_{i}}$, $C=\sum_i^{557}{1}/{\sigma^2_{i}}$,
$\sigma_{i}$ is the 1$\sigma$ uncertainty of the observational data.

Note that the expression
\begin{equation}
\chi^2_{\rm
SNe}=A-\frac{B^2}{C},
\end{equation}
which coincides to Eq. (\ref{chi2SN}) up to a constant \cite{snc}, is used in this paper. The constraint results are the same.

\subsection{Cosmic microwave background}
For CMB, the shift parameters $R$ and $l_a$ are important indicators which are related to the positions of the CMB acoustic peaks. These quantities are
affected by the cosmic expansion history from the decoupling epoch to today.
The shift parameter $R$ is given by \cite{cmbr}
\begin{equation}
R=\Omega_{\mathrm{m0}}^{1/2}\Omega_\mathrm{k}^{-1/2}sinn\bigg[\Omega_\mathrm{k}^{1/2}\int_0^{z_{\ast}}\frac{dz}{E(z)}\bigg],
\end{equation}
where $\Omega_{\mathrm{m0}}=\tilde{\Omega}_m$ (hereafter), since we have defined the generalized dark matter. The redshift $z_{\ast}$ corresponding to the decoupling epoch of photons is given by \cite{cmbz}
\begin{equation}
z_{\ast}=1048[1+0.00124(\Omega_bh^2)^{-0.738}(1+g_{1}(\Omega_{\mathrm{m0}}h^2)^{g_2})],
\end{equation}
with
$g_1=0.0783(\Omega_bh^2)^{-0.238}(1+39.5(\Omega_bh^2)^{-0.763})^{-1}$,
$g_2=0.560(1+21.1(\Omega_bh^2)^{1.81})^{-1}$.
Moreover, the acoustic scale is related to the distance ratio. It is give by
\begin{equation}
l_a=\pi\frac{\Omega_\mathrm{k}^{-1/2}sinn[\Omega_\mathrm{k}^{1/2}\int_0^{z_{\ast}}\frac{dz}{E(z)}]/H_0}{r_s(z_{\ast})},
\end{equation}
where the comoving sound horizon at photo-decoupling epoch is expressed as
\begin{equation}
r_s(z_{\ast})
={H_0}^{-1}\int_{z_{\ast}}^{\infty}c_s(z)/E(z)dz.
\end{equation}

From WMAP 7 measurement, the best-fit values of the data set is given by \cite{cmb}
\begin{eqnarray}
\hspace{-.5cm}\bar{\textbf{P}}_{\rm{CMB}} &=& \left(\begin{array}{c}
{\bar l_a} \\
{\bar R}\\
{\bar z_{\ast}}\end{array}
  \right)=
  \left(\begin{array}{c}
302.09 \pm 0.76\\
1.725\pm 0.018\\
1091.3 \pm 0.91 \end{array}
  \right).
 \end{eqnarray}
 where $\Delta\bf{P_{\mathrm{CMB}}} =
\bf{P_{\mathrm{CMB}}}-\bf{\bar{P}_{\mathrm{CMB}}}$, and the
corresponding inverse  covariance matrix is
\begin{eqnarray}
\hspace{-.5cm} {\bf C_{\mathrm{CMB}}}^{-1}=\left(
\begin{array}{ccc}
2.305 &29.698 &-1.333\\
29.698 &6825.270 &-113.180\\
-1.333 &-113.180 &3.414
\end{array}
\right).
\end{eqnarray}

With these, we can calculate the $\chi^2$ by \cite{cmb}
\begin{eqnarray}
\chi^2_{\mathrm{CMB}}=\Delta
\textbf{P}_{\mathrm{CMB}}^\mathrm{T}{\bf
C_{\mathrm{CMB}}}^{-1}\Delta\textbf{P}_{\mathrm{CMB}}.
\end{eqnarray}

\subsection{Baryon acoustic oscillation}
The Baryon Acoustic Oscillations we use measure the distance-redshift relation at two redshifts. One is in the clustering of the combined 2dFGRS and SDSS main galaxy samples at $z=0.2$,
the other is in the clustering of the SDSS luminous red galaxies at $z=0.35$. Calculated from these samples, the observed scale
of the BAO are analyzed to constrain the form of the distance scale.
The distance scale of BAO is defined as \cite{DV}
\begin{equation}
D_V(z)=c\left(\frac{z}{\Omega_k
H(z)}\mathrm{sinn}^2[\sqrt{|\Omega_k|}\int_0^z\frac{dz'}{H(z')}]\right)^{1/3}.
\end{equation}

The peak positions of the BAO depend on the ratio of $D_v(z)$ to the sound horizon size at the drag epoch.
The  redshift $z_d$ corresponding to the drag epoch when baryons were released from photos is given by
\begin{equation}
z_{d}=\frac{1291(\Omega_{\mathrm{m0}}h^2)^{0.251}}{[1+0.659(\Omega_\mathrm{m0}h^2)^{0.828}]}[(1+b_{1}(\Omega_{b}h^2)^{b_2})],
\end{equation}
where
$b_1=0.313(\Omega_{\mathrm{m0}}h^2)^{-0.419}[1+0.607(\Omega_{\mathrm{m0}}h^2)^{0.674}]^{-1}$
and $b_2=0.238(\Omega_{\mathrm{m0}}h^2)^{0.223}$ \cite{baozd}.
In this paper, we choose the measurement of the distance radio ($d_z$) at $z=0.2$ and $z=0.35$ \cite{baodz}.
This quantity is given by
\begin{equation}
d_z=\frac{r_s(z_d)}{D_V(z)},
\end{equation}
where $r_s(z_d)$ is the comoving sound horizon.
From the SDSS data release 7 (DR7) galaxy sample, the best-fit values
of the data set ($d_{0.2}$, $d_{0.35}$) \cite{baodz}
\begin{eqnarray}
\hspace{-.5cm}\bar{\bf{P}}_{\rm matrix} &=& \left(\begin{array}{c}
{\bar d_{0.2}} \\
{\bar d_{0.35}}\\
\end{array}
  \right)=
  \left(\begin{array}{c}
0.1905\pm0.0061\\
0.1097\pm0.0036\\
\end{array}
  \right),
 \end{eqnarray}
where  the corresponding inverse covariance matrix is
\begin{eqnarray}
\hspace{-.5cm} {\bf C_{\rm matrix}}^{-1}=\left(
\begin{array}{ccc}
30124& -17227\\
-17227& 86977\\
\end{array}
\right).
\end{eqnarray}

The $\chi^2$ value of this BAO observation from SDSS DR7 can be
calculated as
\begin{eqnarray}
\chi^2_{\rm BAO}=\Delta
\textbf{P}_{\rm matrix}^\mathrm{T}{\bf
C_{\rm matrix}}^{-1}\Delta\textbf{P}_{\rm matrix}.
\end{eqnarray}
where $\Delta\bf{P_{\mathrm{matrix}}} =
\bf{P_{\mathrm{matrix}}}-\bf{\bar{P}_{\mathrm{matrix}}}$.

\begin{figure}[t]
  \includegraphics[width=20cm]{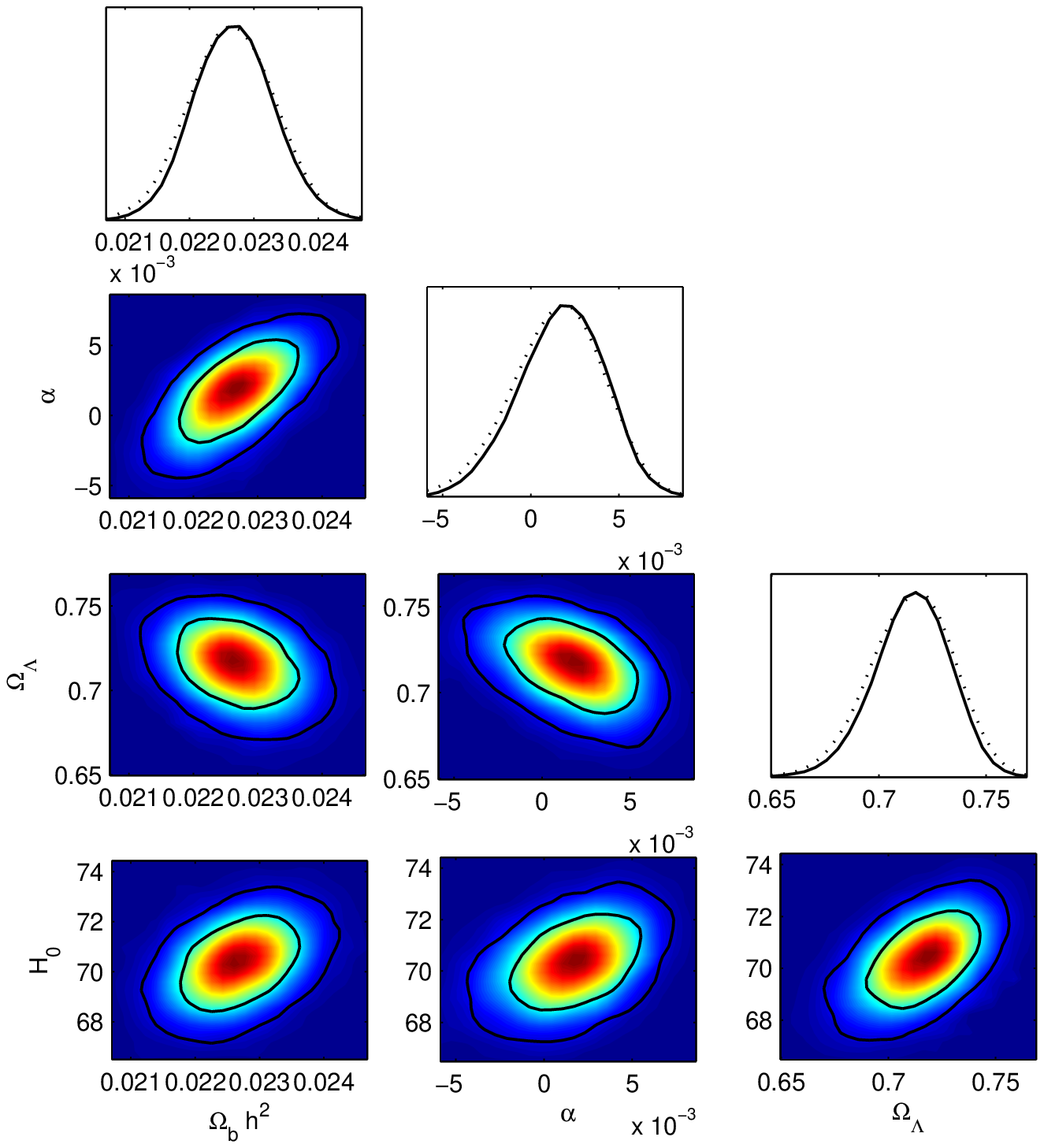}
 \\
  \caption{The 2-D regions and 1-D marginalized distribution with the
1-$\sigma$ and 2-$\sigma$ contours of parameters
$\Omega_bh^2$, $\alpha$,
 $\Omega_\Lambda$ in a flat universe, for the data sets
$H(z)$+CMB+BAO.}\label{figure1}
\end{figure}

\begin{figure}[t]
  \includegraphics[width=20cm]{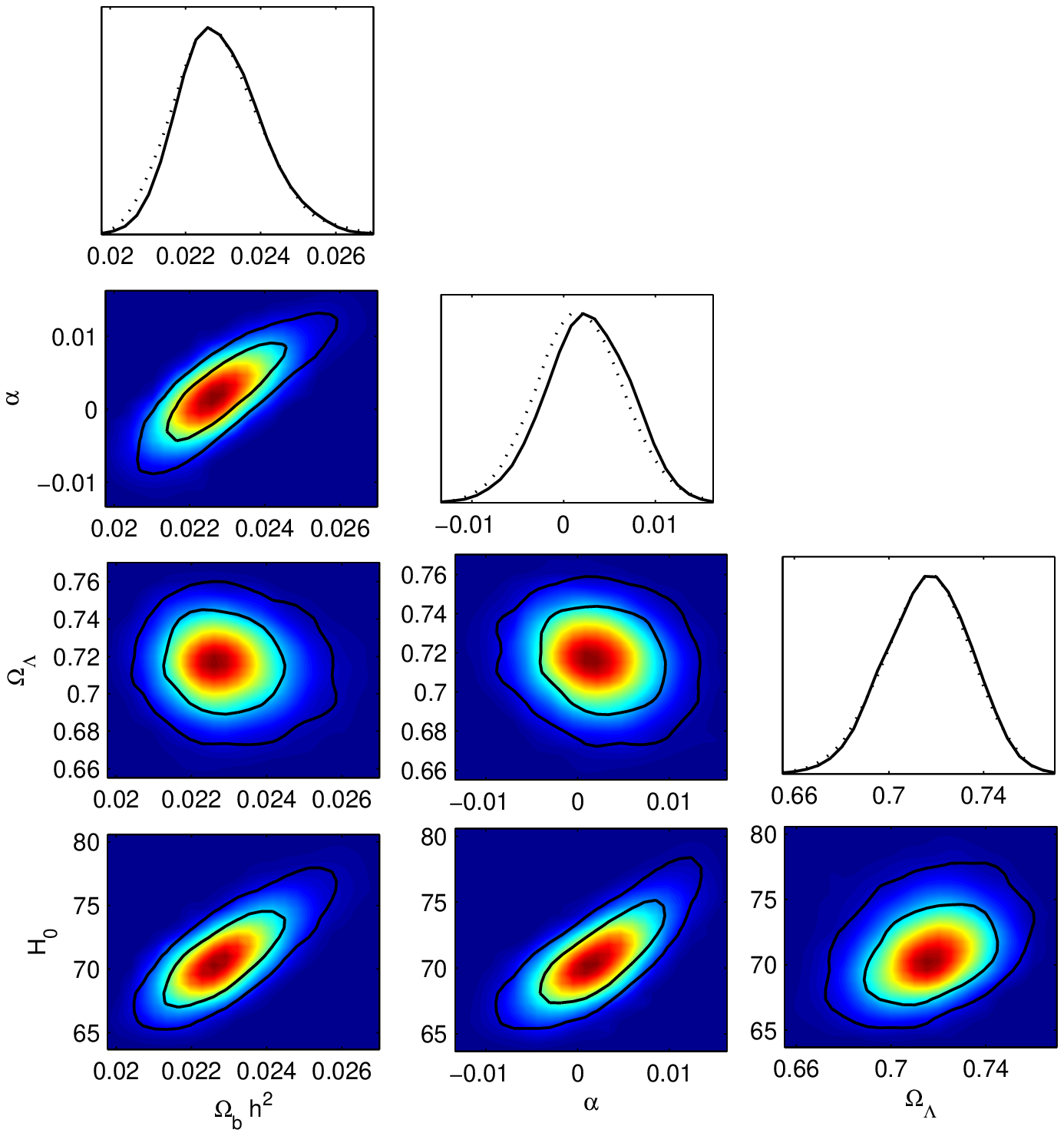}
 \\
  \caption{The 2-D regions and 1-D marginalized distribution with the
1-$\sigma$ and 2-$\sigma$ contours of parameters
$\Omega_bh^2$, $\alpha$,
 $\Omega_\Lambda$ in a flat universe, for the data sets
SNe+CMB+BAO.}\label{figure2}
\end{figure}

\begin{figure}[t]
  \includegraphics[width=20cm]{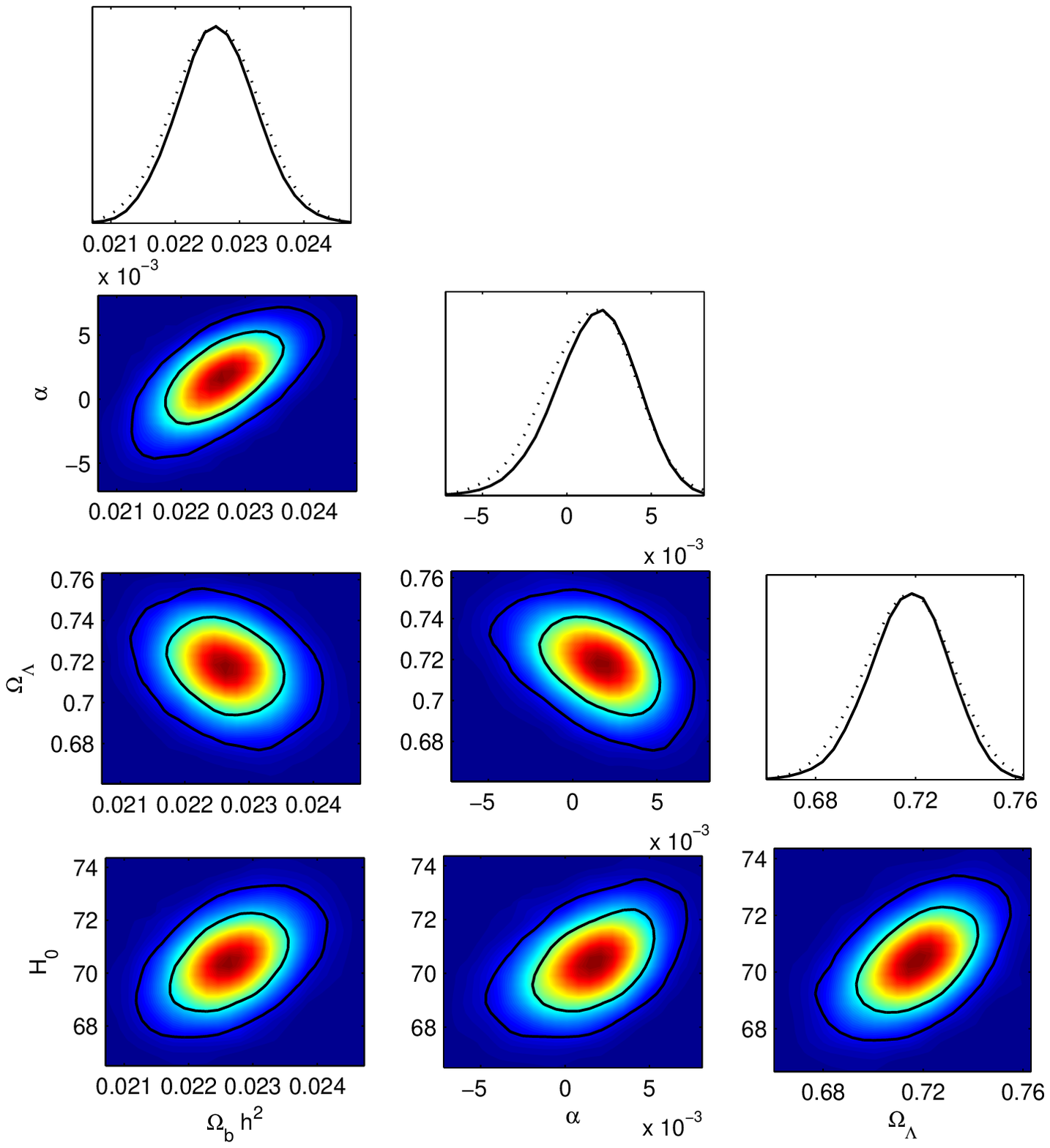}
 \\
  \caption{The 2-D regions and 1-D marginalized distribution with the
1-$\sigma$ and 2-$\sigma$ contours of parameters
$\Omega_bh^2$, $\alpha$,
 $\Omega_\Lambda$ in a flat universe, for the data sets
$H(z)$+SNe+CMB+BAO.}\label{figure3}
\end{figure}

\begin{figure}[t]
  \includegraphics[width=20cm]{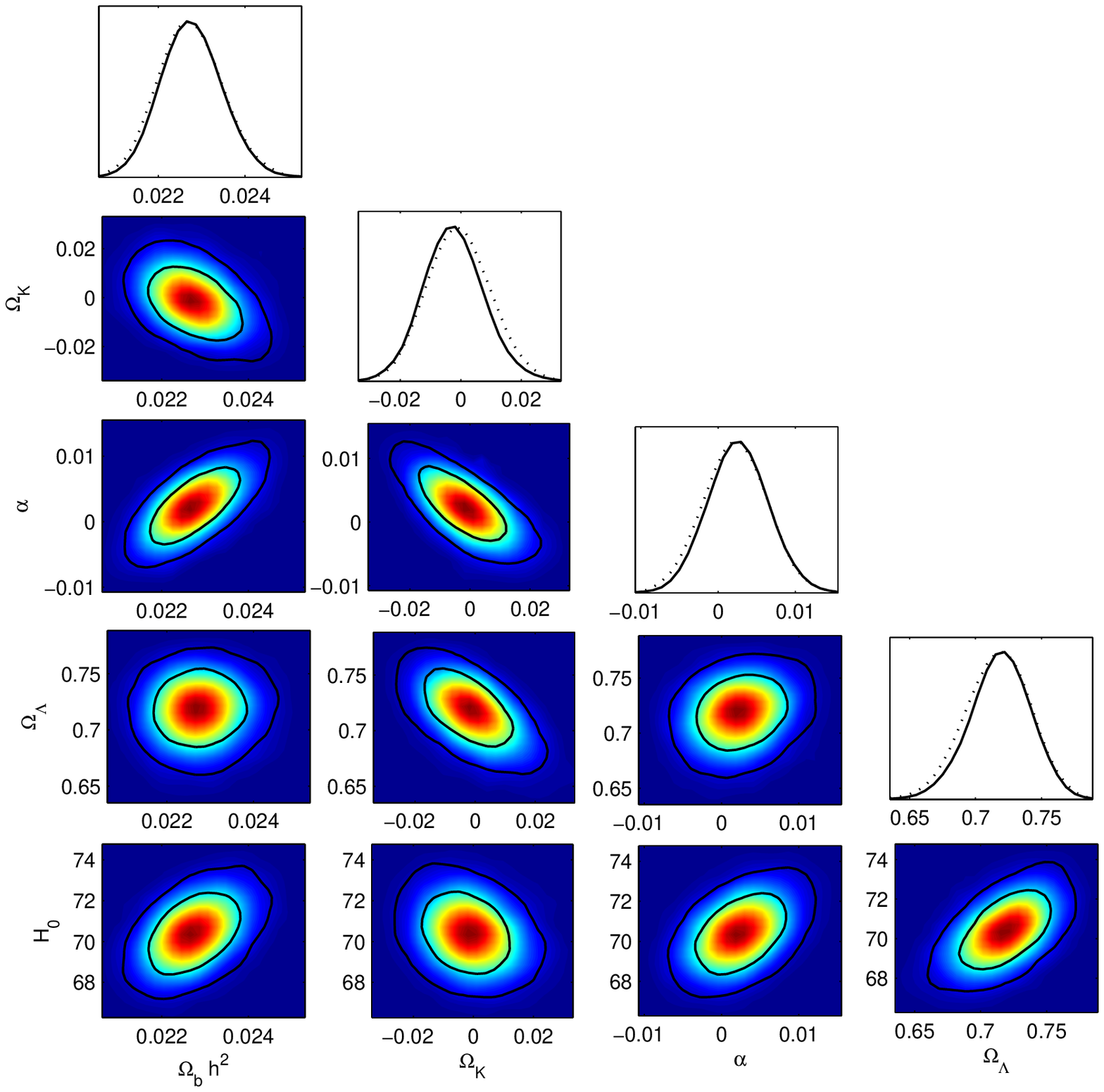}
 \\
  \caption{The 2-D regions and 1-D marginalized distribution with the
1-$\sigma$ and 2-$\sigma$ contours of parameters
$\Omega_bh^2$, $\alpha$,
 $\Omega_\Lambda$, $\Omega_k$ in a non-flat universe, for the data sets
$H(z)$+CMB+BAO.}\label{figure4}
\end{figure}

\begin{figure}[t]
  \includegraphics[width=20cm]{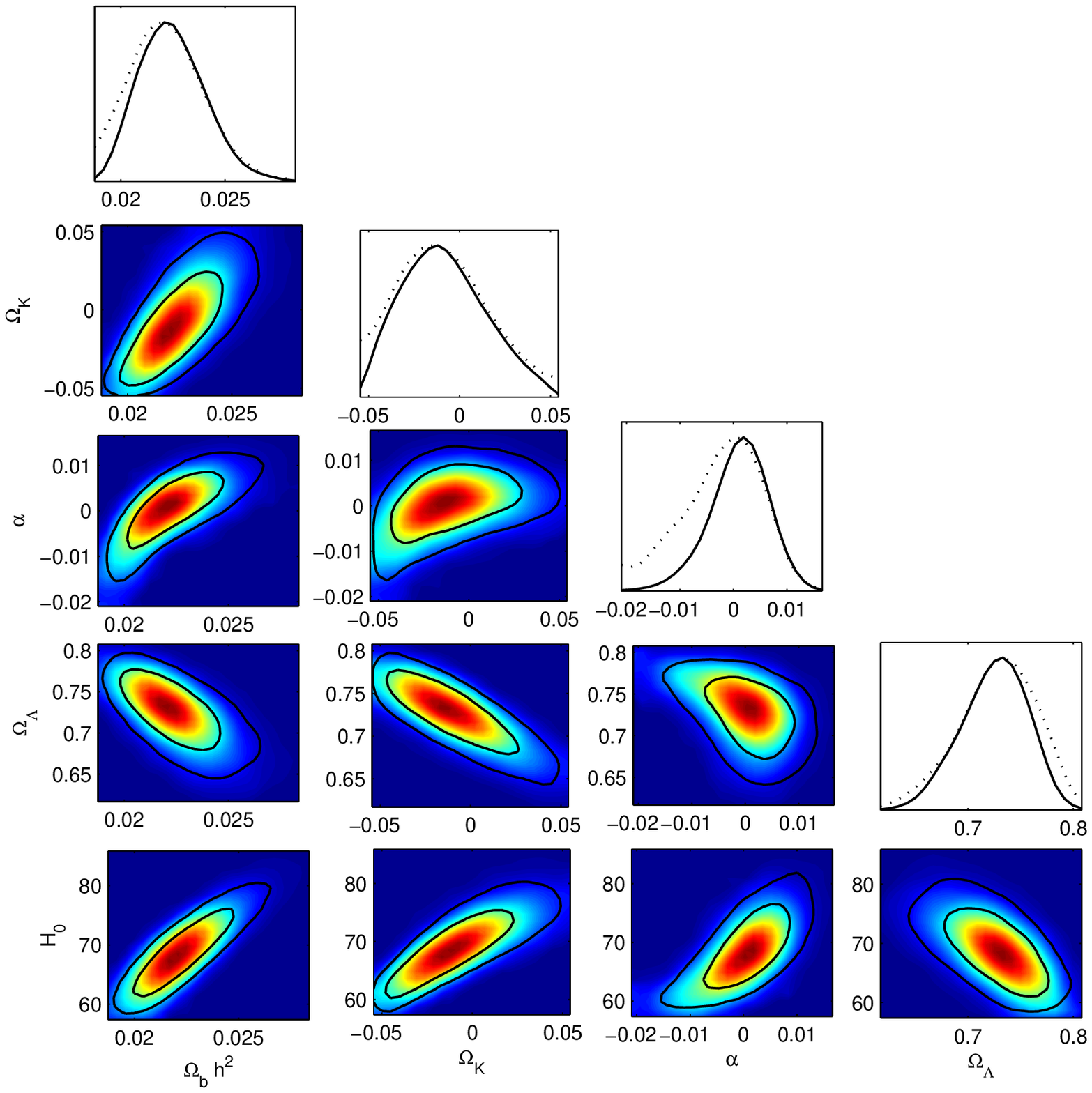}
 \\
  \caption{The 2-D regions and 1-D marginalized distribution with the
1-$\sigma$ and 2-$\sigma$ contours of parameters
$\Omega_bh^2$, $\alpha$,
 $\Omega_\Lambda$, $\Omega_k$ in a non-flat universe, for the data sets
SNe+CMB+BAO.}\label{figure5}
\end{figure}

\begin{figure}[t]
  \includegraphics[width=20cm]{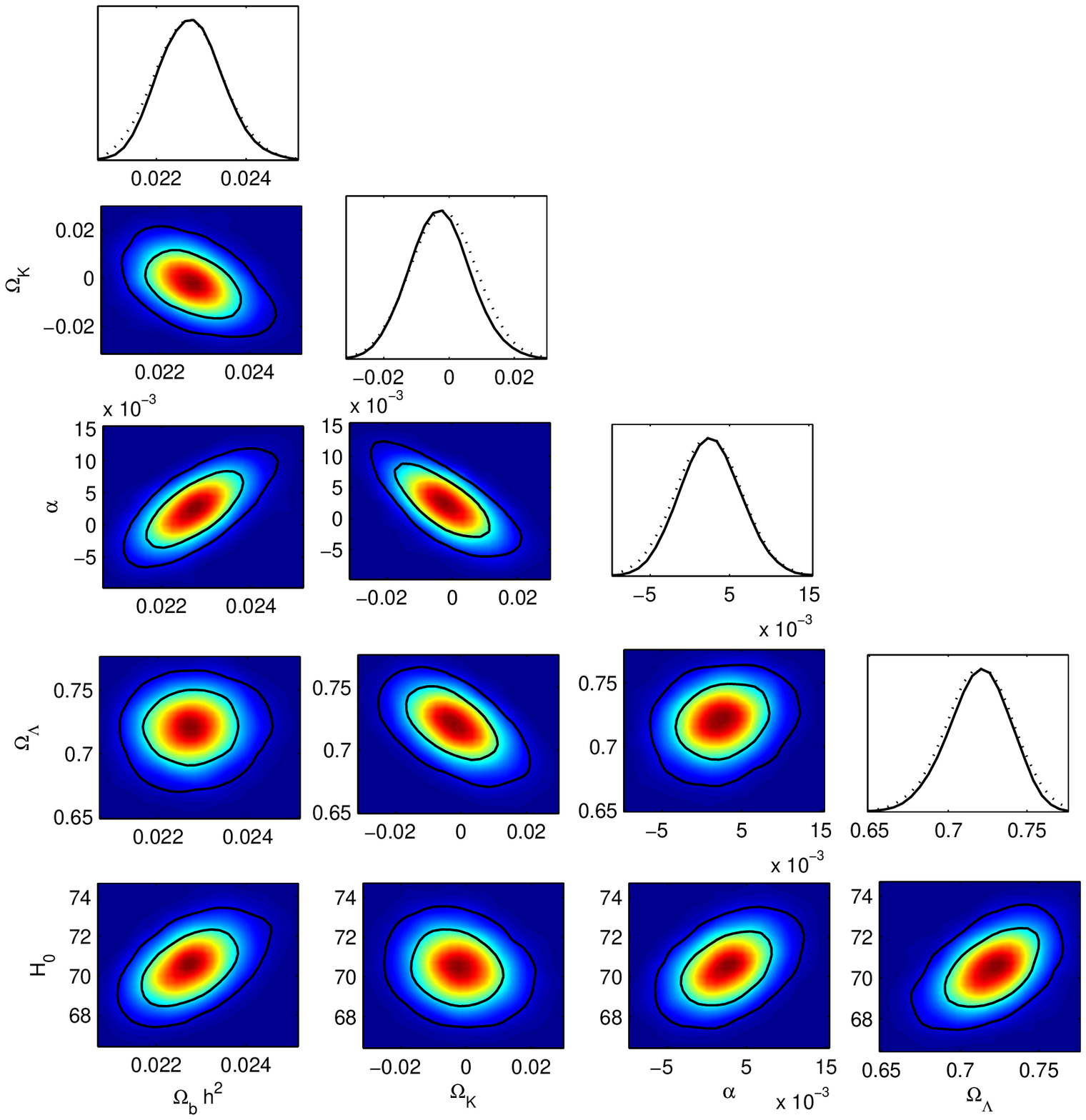}
 \\
  \caption{The 2-D regions and 1-D marginalized distribution with the
1-$\sigma$ and 2-$\sigma$ contours of parameters
$\Omega_bh^2$, $\alpha$,
 $\Omega_\Lambda$, $\Omega_k$ in a non-flat universe, for the data sets
$H(z)$+SNe+CMB+BAO.}\label{figure6}
\end{figure}

\begin{figure}[t]
  \includegraphics[width=16cm]{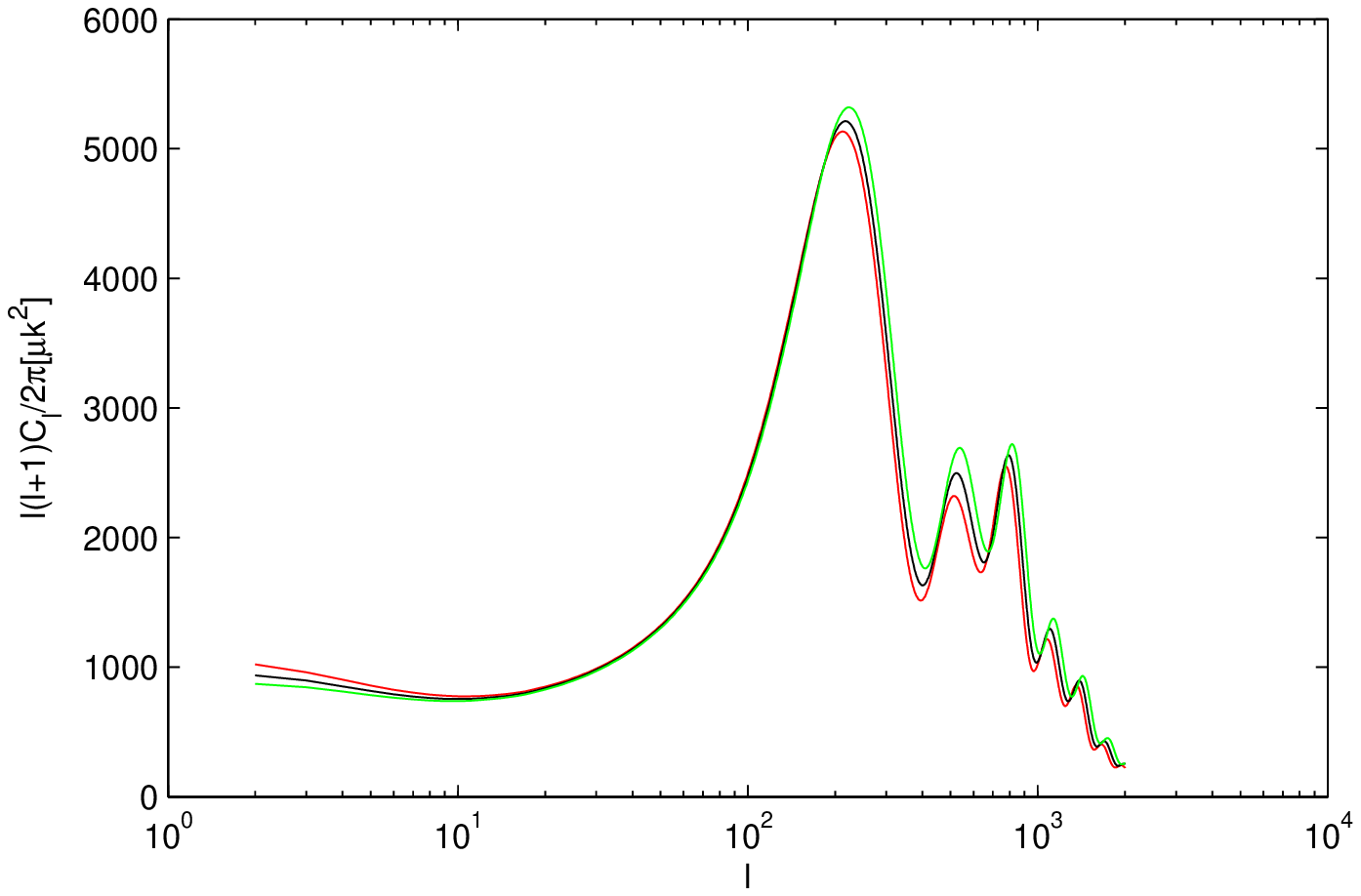}
 \\
  \caption{The theoretical CMB temperature angular power spectrum of UDM model. $\alpha=0.01 (red), 0 (black) and -0.01 (green)$. $\Omega_bh^2=0.022, \Omega_\Lambda=0.71, H_0=70$.}\label{figure7}
\end{figure}

\section{$\text{CONSTRAINT RESULTS}$}
The statistics we use to constrain parameters is the usual maximum likelihood method of $\chi^{2}$ fitting with the Markov Chain
Monte Carlo (MCMC) method. Our code is based on CosmoMCMC \cite{mcmc}. The corresponding convergence of the chains R-1 is set to be less
than 0.003 which can guarantee the accuracy well. We combine $H(z)$, SNe Ia, CMB and BAO data to give a global fitting on determining
the parameters of UDM model. The total $\chi^{2}$ is given by
\begin{equation}
\chi^2_{tot}=\chi^2_{H}+\chi^2_{\rm SNe}+\chi^2_{\rm CMB}+\chi^2_{\rm BAO}.
\end{equation}

Fist, we consider a flat universe. In order to embody the $H(z)$ data, we choose three combinations of data sets and show the 1-D probability distribution of each parameter ($\Omega_bh^2$, $\Omega_\Lambda$, $\alpha$) and 2-D plots for parameters between each other. Fitting
results from the joint data of $H(z)$+CMB+BAO are given in Fig 1, the best-value for $\alpha$ is
$\alpha$=$1.61_{-4.58}^{+3.99}(1\sigma)_{-6.92}^{+5.65}(2\sigma)(\times10^{-3})$. For comparison, we give the results from
SNe+CMB+BAO in Fig 2. The best-fit value is $\alpha$=$1.81_{-8.10}^{+8.15}(1\sigma)_{-11.69}^{+11.99}(2\sigma)(\times10^{-3})$.
We can see the observational Hubble parameter data gives a more stringent constraint than SNe Ia in this model.
In Fig 3, the total combinations of them result in $\alpha$=$1.72_{-4.79}^{+3.92}(1\sigma)_{-7.30}^{+5.47}(2\sigma)(\times10^{-3})$. From the
result, the parameter $\alpha$ which stands for the speed of sound of the unified dark sector seems to be vanishing or slightly larger than 0. So $\Lambda$CDM model is strongly supported. We further consider a non-flat universe with curvature. For $H(z)$+CMB+BAO in Fig 4, the
results are $\alpha$=$2.21_{-7.60}^{+7.98}(1\sigma)_{-10.91}^{+11.06}(2\sigma)(\times10^{-3})$, $\Omega_k$=$-0.208_{-1.88}^{+2.27}(1\sigma)_{-2.60}^{+3.12}(2\sigma)(\times10^{-2})$. Using SNe+CMB+BAO in Fig 5,
$\alpha$=$0.662_{-22.52}^{+9.80}(1\sigma)_{-37.33}^{+13.68}(2\sigma)(\times10^{-3})$,
$\Omega_k$=$-1.41_{-4.83}^{+6.35}(1\sigma)_{-6.46}^{+9.38}(2\sigma)(\times10^{-2})$. This results are consist with the situation in a flat universe. The $H(z)$ data also gives a much more stringent constraint on $\Omega_k$. The total joint data in Fig 6 give
$\alpha$=$2.42_{-7.75}^{+7.87}(1\sigma)_{-10.29}^{+11.00}(2\sigma)(\times10^{-3})$,
$\Omega_k$=$-0.194_{-1.85}^{+2.02}(1\sigma)_{-2.57}^{+2.75}(2\sigma)(\times10^{-2})$, which suggest both the speed of sound and the curvature are very little.
Our results are much more stringent than previous work \cite{constraint}. \textbf{Our results seem to be consistent with
$\alpha=0$ using the data sets we select. Does that mean there is no need to introduce a new parameter? We think although the UDM model is contrived, the sound speed of the unified dark matter is related to the microphysics of the dark component while equation of state does not. The sound speed leads to the inhomogeneities of dark matter, thus affecting CMB and matter power spectra. With the development of technique, more information in CMB, matter power spectra and so on may distinguish UDM model from $\Lambda$CDM. Moreover, we assume a constant sound speed, the whole effective value is very small, however, if when we consider the evolving speed of sound, the results may be changed.}
\begin{table*}
 \begin{center}
 \begin{tabular}{|c|c|c|c|} \hline\hline
 & \multicolumn{3}{c|}{Flat Universe}  \\
 \cline{2-4}                 &     $H(z)$+CMB+BAO                  &     SNe+BAO+CMB                       &    $H(z)$+SNe+CMB+BAO          \\ \hline
$\Omega_\Lambda$  \ \ & \ \ $0.717_{-0.0338}^{+0.0285}(1\sigma)_{-0.0506}^{+0.0404}(2\sigma)$ \ \  & \ \ $0.719_{-0.0353}^{+0.0282}(1\sigma)_{-0.0532}^{+0.0412}(2\sigma)$ \ \ & \ \ $0.719_{-0.0305}^{+0.0264}(1\sigma)_{-0.0458}^{+0.0380}(2\sigma)$\ \ \\
$\alpha$  \ \ & \ \ $1.61_{-4.58}^{+3.99}(1\sigma)_{-6.92}^{+5.65}(2\sigma)(\times10^{-3})$ \ \  & \ \ $1.81_{-8.10}^{+8.15}(1\sigma)_{-11.69}^{+11.99}(2\sigma)(\times10^{-3})$ \ \ & \ \ $1.72_{-4.79}^{+3.92}(1\sigma)_{-7.30}^{+5.47}(2\sigma)(\times10^{-3})$\ \ \\
$\Omega_bh^2$                 \ \ & \ \ $0.0226_{-0.0010}^{+0.0012}(1\sigma)_{-0.0015}^{+0.0017}(2\sigma)$ \ \  & \ \ $0.0227_{-0.0016}^{+0.0022}(1\sigma)_{-0.0022}^{+0.0035}(2\sigma)$ \ \ & \ \ $0.0226_{-0.0011}^{+0.0011}(1\sigma)_{-0.0015}^{+0.0016}(2\sigma)$\ \ \\

 \hline\hline
$\chi_{\rm min}^2$              \ \ & \ \ $11.592$\ \                  & \ \ $532.142$\ \                   & \ \ $542.354$\ \      \\
\hline\hline

 \end{tabular}
 \end{center}\label{tab1}
 \caption{The best-fit values of  parameters $\Omega_bh^2$, $\alpha$,
 $\Omega_\Lambda$ for UDM model in a flat universe
 with the 1-$\sigma$ and 2-$\sigma$ uncertainties, as well as $\chi_{\rm min}^2$,
for the data sets $H(z)$+CMB+BAO, SNe+BAO+CMB, and $H(z)$+SNe+BAO+CMB,
respectively. }\label{tab1}
 \end{table*}

\begin{table*}
 \begin{center}
 \begin{tabular}{|c|c|c|c|} \hline\hline
 & \multicolumn{3}{c|}{Non-flat Universe}  \\
 \cline{2-4}                 &     $H(z)$+CMB+BAO                  &     SNe+BAO+CMB                       &    $H(z)$+SNe+CMB+BAO          \\ \hline
$\Omega_\Lambda$  \ \ & \ \ $0.721_{-0.0535}^{+0.0414}(1\sigma)_{-0.0779}^{+0.0549}(2\sigma)$ \ \  & \ \ $0.734_{-0.0815}^{+0.0609}(1\sigma)_{-0.124}^{+0.0799}(2\sigma)$ \ \ & \ \ $0.722_{-0.0447}^{+0.0362}(1\sigma)_{-0.634}^{+0.0479}(2\sigma)$\ \ \\
$\alpha$  \ \ & \ \ $2.21_{-7.60}^{+7.98}(1\sigma)_{-10.91}^{+11.06}(2\sigma)(\times10^{-3})$ \ \  & \ \ $0.662_{-22.52}^{+9.80}(1\sigma)_{-37.33}^{+13.68}(2\sigma)(\times10^{-3})$ \ \ & \ \ $2.42_{-7.75}^{+7.87}(1\sigma)_{-10.29}^{+11.00}(2\sigma)(\times10^{-3})$\ \ \\
$\Omega_bh^2$                 \ \ & \ \ $0.0227_{-0.0013}^{+0.0015}(1\sigma)_{-0.0019}^{+0.0021}(2\sigma)$ \ \  & \ \ $0.0220_{-0.0030}^{+0.0039}(1\sigma)_{-0.0035}^{+0.0056}(2\sigma)$ \ \ & \ \ $0.0227_{-0.0014}^{+0.0015}(1\sigma)_{-0.0018}^{+0.0021}(2\sigma)$\ \ \\
$\Omega_k$                 \ \ & \ \ $-0.208_{-1.88}^{+2.27}(1\sigma)_{-2.60}^{+3.12}(2\sigma)(\times10^{-2})$ \ \  & \ \ $-1.41_{-4.83}^{+6.35}(1\sigma)_{-6.46}^{+9.38}(2\sigma)(\times10^{-2})$ \ \ & \ \ $-0.194_{-1.85}^{+2.02}(1\sigma)_{-2.57}^{+2.75}(2\sigma)(\times10^{-2})$\ \ \\
 \hline\hline
$\chi_{\rm min}^2$              \ \ & \ \ $11.578$\ \                  & \ \ $531.870$\ \                   & \ \ $542.309$\ \      \\
\hline\hline

 \end{tabular}
 \end{center}\label{tab2}
 \caption{The best-fit values of  parameters $\Omega_bh^2$, $\alpha$,
 $\Omega_\Lambda$, $\Omega_k$ for UDM model in a non-flat universe
 with the 1-$\sigma$ and 2-$\sigma$ uncertainties, as well as $\chi_{\rm min}^2$,
for the data sets $H(z)$+CMB+BAO, SNe+BAO+CMB, and $H(z)$+SNe+BAO+CMB,
respectively. }\label{tab2}
 \end{table*}

\section{$\text{CONCLUSIONS}$}
In this paper, we constrain the unified dark matter with constant speed of sound from the latest data: the newly revised 15 Hubble data, type Ia supernovae (SNe Ia) from Union2 set,
baryonic acoustic oscillation (BAO) observation from the spectroscopic Sloan Digital Sky
Survey (SDSS) data release 7 (DR7) galaxy sample, as well as the cosmic microwave
background (CMB) observation from the 7-year Wilkinson Microwave Anisotropy Probe (WMAP7) results.
We consider two case. In a flat universe, the constraint results are
$\Omega_\Lambda$=$0.719_{-0.0305}^{+0.0264}(1\sigma)_{-0.0458}^{+0.0380}(2\sigma)$,
$\alpha$=$1.72_{-4.79}^{+3.92}(1\sigma)_{-7.30}^{+5.47}(2\sigma)(\times10^{-3})$,
$\Omega_bh^2$=$0.0226_{-0.0011}^{+0.0011}(1\sigma)_{-0.0015}^{+0.0016}(2\sigma)$.
In a non-flat universe, the constraint results are
$\Omega_\Lambda$=$0.722_{-0.0447}^{+0.0362}(1\sigma)_{-0.0634}^{+0.0479}(2\sigma)$,
$\alpha$=$0.242_{-0.775}^{+0.787}(1\sigma)_{-1.03}^{+1.10}(2\sigma)(\times10^{-2})$,
$\Omega_bh^2$=$0.0227_{-0.0014}^{+0.0015}(1\sigma)_{-0.0018}^{+0.0021}(2\sigma)$,
$\Omega_k$=$-0.194_{-1.85}^{+2.02}(1\sigma)_{-2.57}^{+2.75}(2\sigma)(\times10^{-2})$.
This constraint results are much stringent than previous work \cite{constraint}, which suggest both the speed of sound of the UDM and
the curvature are very little and a flat $\Lambda$CDM model is still a good choice. In order to know the effects of the observational
Hubble data, we also consider the combinations $H(z)$+CMB+BAO and SNe+BAO+CMB. From the results, the $H(z)$ data gives a better constraint
especially for $\alpha$, $\Omega_k$ and $\Omega_bh^2$ than SNe Ia in this model. We think the $H(z)$ data will paly an important role in
cosmology in the future \cite{hzfuture}.

\textbf{\ Acknowledgments } We are grateful to Lixin Xu  for introducing the MCMC method. We also thank Hao Wang and Jun Wang for helpful discussions. This work was supported by the National Natural Science Foundation of
China under the Distinguished Young Scholar Grant 10825313 and Grant 11073005,
the Ministry of Science and Technology national basic science Program (Project 973)
under Grant No.2012CB821804, the Fundamental Research Funds for the Central
Universities and Scientific Research Foundation of Beijing Normal University.

\end{document}